# Superconductivity in a Bulk Moiré Superlattice Material

*Subham Naik[a], Paul Monson[a], Susanta Manna[b], Prabuddha Kant Mishra[c], Soumyojit Chatterjee[d], Sandip Kuila[e], Partha Pratim Jana[e], M. B. Sreedhara[b], Rahul Sharma[d]* and Gohil S. Thakur[a]**

[a]*Department of Chemical Sciences, Indian Institute of Science Education and Research, Berhampur-760003, Odisha, India*
[b]*Solid State and Structural Chemistry Unit, Indian Institute of Science, Bengaluru, 560012, India*
[c]*Central Advanced Instrument Facility, Indian Institute of Science Education and Research, Berhampur-760003, Odisha, India*
[d]*Department of Physical Sciences, Indian Institute of Science Education and Research, Berhampur-760003, Odisha, India*
[e]*Department of Chemistry, Indian Institute of Technology, Kharagpur*, 721302

## Abstract

**Moiré materials provide a versatile platform realizing emergent electronic states arising from enhanced correlation due to flat bands.[1] A variety of phenomena including superconductivity, low-dimensional ferromagnetism, Mott insulating phase, and topological phenomena have been reported in such systems.[2–13] To date, moiré phenomena have been predominantly explored in artificially assembled low-dimensional van der Waals structures, where relative twist and lattice alignment are controlled during device fabrication[2–4,6,10,12–14]. Recently, intrinsically grown bulk moiré crystals have emerged as a complementary materials platform, in which lattice mismatch between constituent layers generates a coherent moiré superlattice throughout the bulk crystal[15]. Here, we report the evidence of bulk superconductivity in single crystals of a recently reported bulk Moiré material $(Sr_6TaS_8)_{1+\delta}(TaS_2)_8$ at ambient pressure. The material exhibits a bulk superconducting transition at $T_c \simeq$ 2.5 K, evidenced consistently by electrical transport, magnetic susceptibility, and heat-capacity measurements on single crystals. Structural characterization by X-ray diffraction and atomic-resolution electron microscopy establishes the composite, incommensurate crystal structure and scanning tunnelling microscopy reveals a long-wavelength surface modulation due to the incommensurate structure. The observation of bulk superconductivity establishes superconductivity as an emergent phase in this intrinsically synthesized moiré material and highlights bulk moiré crystals as a promising platform for investigating correlated quantum phenomena beyond artificially assembled two-dimensional structures.**

## Main

Exotic electronic phenomena can emerge in materials composed of periodically modulated atomic layers, commonly referred to as superlattice materials[1]. Moiré materials constitute a particularly versatile class of such superlattice systems, in which a small lattice mismatch and/or relative rotational misalignment between constituent layers produces a long-wavelength spatial modulation. The resulting moiré potential can reconstruct the low-energy electronic structure and provide a platform for realizing correlated, magnetic, topological and optical phenomena distinct from those of its constituents.[4,10,12,16–25] A central attraction of moiré systems is that their emergent properties can be tuned through geometric parameters such as twist angle or lattice mismatch, offering a route to engineer quantum phases without necessarily changing the chemical composition. This approach has led to the observation of correlated insulating states and superconductivity in twisted graphene systems, as well as correlated and magnetic states in transition-metal dichalcogenide (TMD) moiré heterostructures.[5,6,8,9,26–30] More recently, superconductivity has also been observed in twisted bilayer $WSe_2$, demonstrating that superconducting phases are not restricted to graphene-based moiré systems.[5,31]

Moiré physics has been predominantly explored in artificially assembled van der Waals heterostructures, in which atomically thin layers are stacked with a controlled relative orientation, or in epitaxially grown heterostructures.[6,9,32] Such platforms provide exceptional control over carrier density, displacement field and twist angle, but require the fabrication and preservation of atomically aligned interfaces and are generally realized as individual nanoscale or microscale structures. An alternative route is to generate the moiré structure intrinsically during crystal growth, thereby combining a long-wavelength moiré superlattice with the macroscopic dimensions and robustness of a crystalline material. Such an approach is particularly attractive for studying emergent phases, as macroscopic crystals provide access to complementary transport, magnetic, thermodynamic, and high-field measurements that can be challenging in artificially assembled two-dimensional heterostructures.

A breakthrough development in this direction was the recent discovery of a family of intrinsically grown bulk moiré crystals $(Sr_6TaS_8)_{1+\delta}(TaS_2)_8$ in which lattice mismatches between alternating $TaS_2$ and $Sr_6TaS_8$ layers generate incommensurate moiré superlattices that are coherent throughout the crystals.[15] Importantly, the moiré features can be tuned via the synthesis conditions without changing the chemical composition, providing a materials-based route to engineering moiré electronic structures. Quantum-oscillation measurements revealed a complex Fermiology that can be naturally understood in terms of a higher-dimensional description of the underlying incommensurate crystal structure. These crystals are also easily exfoliable, providing a potential connection between intrinsically grown bulk moiré materials and low-dimensional device architectures.[15] More broadly, intrinsic synthesis of moiré crystals offers a complementary route to artificially assembled heterostructures and potentially provides a scalable approach to producing tunable macroscopic moiré materials.

The unusual electronic structure associated with incommensurate moiré systems also raises the possibility of unconventional ordered states beyond those realized in conventional periodic systems.[33] Experimental realization of superconductivity in such an intrinsically synthesized bulk moiré crystal would therefore provide an opportunity to investigate superconductivity in a moiré environment using macroscopic single crystals and complementary transport, magnetic and thermodynamic probes.

Here, we report bulk superconductivity below $T_c \simeq 2.5$ K in single crystals of $(Sr_6TaS_8)_{1+\delta}(TaS_2)_8$. The superconducting transition is consistently observed in electrical transport, magnetic susceptibility and heat-capacity measurements, establishing a bulk superconducting state in this recently discovered bulk moiré material. Moreover, we provide a direct real-space visualization of the long wavelength modulation through scanning tunnelling microscopy. Our results establish superconductivity as an emergent ordered state in an intrinsically synthesized bulk moiré crystal and provide a platform for investigating the interplay between moiré electronic structure and superconductivity in a macroscopic quantum material.

## Evidence of Moiré Superlattice

Fig. 1 shows a reconstructed reciprocal-space $hk4$ section, obtained from the single-crystal X-ray diffraction measurements, demonstrating the three-dimensional crystalline order of the structure. The strong reflections marked by red circles correspond to the hexagonal $TaS_2$ sublattice (Fig. 1). Similar well-defined arrays of intense diffraction spots is observed in all *hkl* sections, from $l = -4$ to $+4$ (Supplementary Information). Their positions remain well defined with changing $l$, confirming that these reflections originate from the underlying $TaS_2$ lattice rather than from random scattering or diffuse features. In addition to the principal Bragg reflections, a number of weaker satellite spots (blue circles in Fig. 1b) are visible around the strong reflections. These weak reflections do not belong to the basic $TaS_2$ reciprocal lattice; instead, they arise from the additional periodicity generated by the contribution from both sub-lattices ($TaS_2$ and $Sr_6TaS_8$) (Fig. 1).

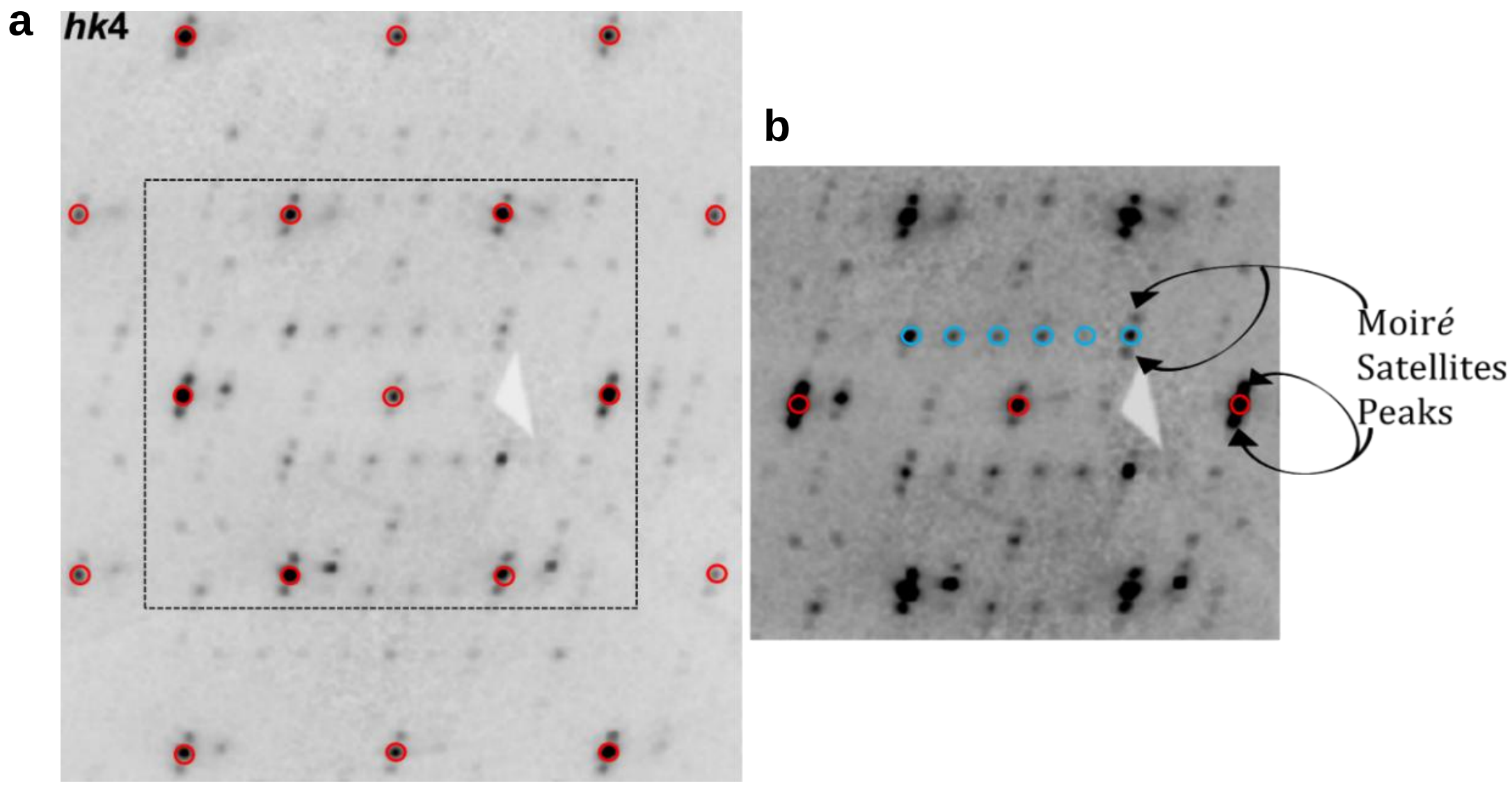

**Fig. 1| Single Crystal X-ray diffraction study. a,** Reciprocal reconstruction image of $hk4$. **b,** Zoomed-in plot of dashed box showing $q$ vector satellite peaks appearing near dominant Bragg reflections of $TaS_2$ (red circle) and $Sr_6TaS_8$ (blue circles) layers.

Consequently, the reciprocal-space pattern contains two superimposed structural contributions: strong fundamental Bragg reflections from the constituent crystalline sublattices and weaker satellite reflections associated with the incommensurate Moiré modulation. This coexistence is a signature of the composite nature of the structure. For a complete quantitative determination of commensurate Bragg plane order ($n$) and angle of the Moiré wave vector $q$ with $b$-axis ($\varphi$), wide-angle single crystal X-ray diffraction and further refinement would be needed.

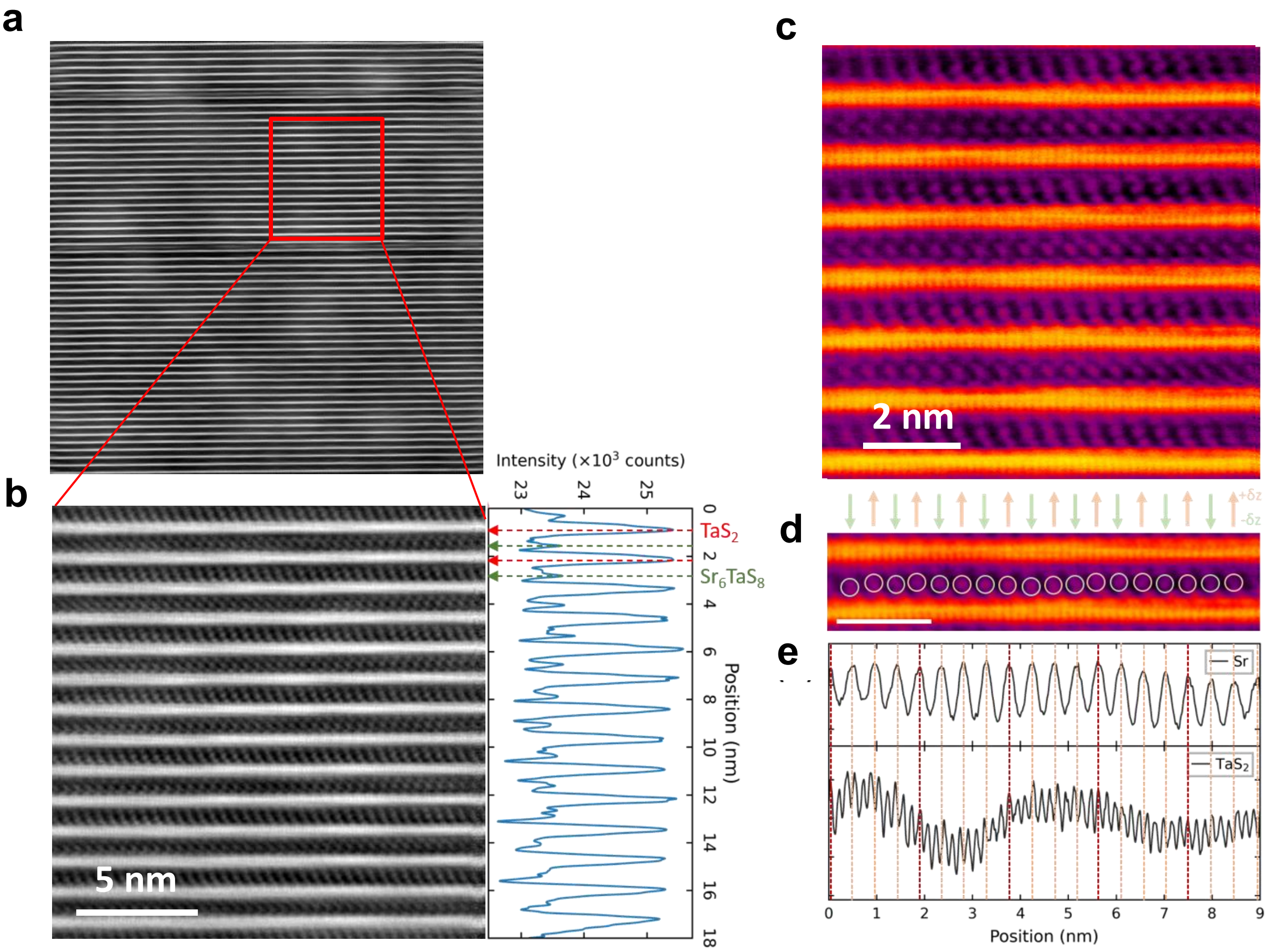

**Fig. 2| High Angle Annular Dark Field-Scanning Transmission Electron Microscope** (HAADF-STEM) image of $(Sr_6TaS_8)_{1+\delta}(TaS_2)_8$ along the [010] zone axis. **a,** Low-magnification HAADF-STEM images of the FIB lamella; The bright atomic columns correspond to the $TaS_2$ sublattices, and the darker layers are $Sr_6TaS_8$. The red box marks the region shown in **b**. **b,** High-resolution HAADF-STEM images and the intensity profile integrated along the $c$-direction. The intensity profile clearly shows the alternating layers of $TaS_2$ and $Sr_6TaS_8$, with the distance between two adjacent $TaS_2$ layers being around 12.58 Å. **c,** Colour High-resolution HAADF-STEM image showing atomically resolved $TaS_2$ and $Sr_6TaS_8$ layers. **d,** HAADF-STEM image of an individual $Sr_6TaS_8$ spacer layer, highlighting the central $TaS_6$ clusters with a characteristic dimerised up-down configuration. The two sublattices are colour-coded according to their HAADF intensities to clearly distinguish the respective layers. **e,** Corresponding HAADF intensity profiles extracted from the individual sublattices. The intensity profiles reveal the incommensurate structural relationship between the $TaS_2$ and $Sr_6TaS_8$ layers.

Fig. 2a shows the low-magnification HAADF-STEM image of the focused-ion beam (FIB) prepared lamella acquired along the [010] zone axis, i.e. the $b$-axis of the crystal, following the $TaS_2$-based notation from ref.[15]. The image reveals a highly periodic long-range ordered layer structure, characterised by alternating bright and dark contrast. As the HAADF intensity strongly depends on the atomic number, the bright layers are assigned to the Ta-rich $TaS_2$ layers and the darker layers to the $Sr_6TaS_8$ spacer layers. A high-magnification HAADF-STEM image in Fig. 2b, together with the intensity profile integrated along the *c*-direction, shows a regular sequence of strong and weak maxima arising from the $TaS_2$ and $Sr_6TaS_8$ layers, respectively. The pronounced periodicity of these intensity maxima provides direct real-space evidence for the regular stacking of the two structurally distinct subsystems, revealing the periodic structure along the *c*-direction. The separation between successive $TaS_2$ layers is approximately 12.58 Å, matching the *c*-lattice parameter of $(Sr_6TaS_8)_{1+\delta}(TaS_2)_8$, consistent with the recent report.[15] The alternating stacking is also confirmed in reciprocal space; the selected area electron diffraction (SAED) shows (000*l*) reflections with $c$ = 12.81 Å and reflections at every integer l, including positions that are systematically absent in 2*H*-$TaS_2$ (Supplementary Information). This confirms that each stacking period contains a single layer of $TaS_2$ and alternates with a $Sr_6TaS_8$ layer. The strongest in-plane reflections lie around the $TaS_2$ $\{11\bar{2}0\}$ spacing (1.58 Å), identifying the zone axis as $TaS_2$ <120>-type. In addition, weak rows appear between the 000l row and the $TaS_2$ $\{11\bar{2}0\}$ group, near one-third (indicated by the blue arrow) and two-thirds (indicated by the orange arrow) of the $TaS_2$ $\{11\bar{2}0\}$ reciprocal spacing; these rows have no counterpart in $TaS_2$. Such a periodicity is foreign to $TaS_2$ and is attributed to the $Sr_6TaS_8$ spacer lattice and to mixed reflections of the two subsystems.

The atomically resolved HAADF-STEM images in Fig. 2c-d show the in-plane structure of both subsystems. In this projection, the horizontal direction is the *a*-axis ([210]), the direction along which the $TaS_2$ and $Sr_6TaS_8$ lattices are incommensurate (ref. [15]). The HAADF intensity profiles extracted along the *a*-direction from the $TaS_2$ layer and from the adjacent $Sr_6TaS_8$ layers (Fig. 2e) show distinct periodicities of the two constituent subsystems. Their relative registry progressively deviates along the *a*-direction, evidently demonstrating the incommensurate relationship between the $TaS_2$ and $Sr_6TaS_8$ lattices. The orange dotted lines in Fig. 2e highlight the relative positional mismatch between the Ta atomic columns of the two subsystems. Notably, the Ta atomic periodicity of the $TaS_2$ layer extends over 12 orders, whereas the corresponding periodicity of the $Sr_6TaS_8$ layer extends over 5 orders, providing direct real-space evidence of the lattice mismatch between the two subsystems. The atomic-scale structure of an individual $Sr_6TaS_8$ spacer layer is further resolved in the HAADF-STEM image shown in Fig. 2d. The central $TaS_6$ clusters exhibit a characteristic dimerised up-down displacement configuration, highlighted by the green and orange arrows and corresponding circles. This periodic displacement of the $TaS_6$ clusters produces a pronounced structural modulation along the *z*-direction, providing direct visualisation of the moiré-like modulation within the $Sr_6TaS_8$ spacer layer.

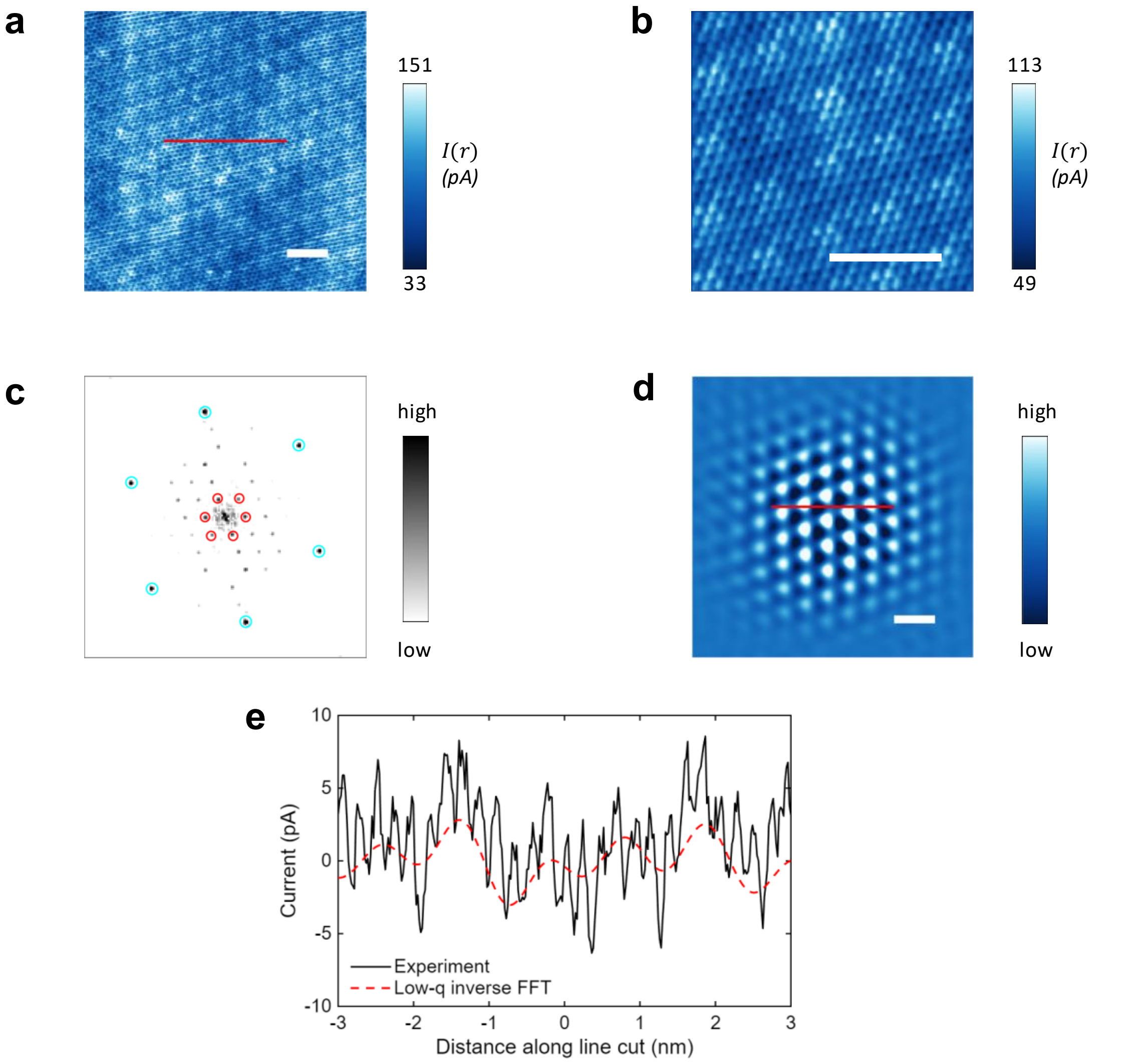

**Fig. 3| Atomic-scale structure and long-wavelength modulation revealed by STM.** **a,** Atomically resolved STM tunnelling-current map acquired simultaneously with the $Z$-channel during constant-current imaging at $V_s = 300$ mV and $I_s = 100$ pA, showing the atomic-scale structure together with a longer-wavelength modulation. The red line indicates the trajectory of the representative line cut shown in **e**, and the white bar denotes the 2 nm scalebar. **b,** STM tunnelling current map $I(r)$ recorded during constant current imaging at $V_s = 400$ mV and $I_s = 80$ pA in a separate smaller FOV, highlighting the long-wavelength modulation superimposed on the atomic-scale structure. Scalebar is 2 nm. **c,** Fourier transform of the STM current map, showing the atomic-scale Fourier components (cyan circles) together with additional low-$q$ modulation peaks (red circles). **d,** Real-space reconstruction obtained by inverse Fourier transforming the selected observed low-$q$ Fourier components marked by red circles in **c**, revealing the corresponding long-wavelength modulation. **e,** Representative line cut along the trajectory marked in **a**, showing the spatial variation of the experimental tunnelling current (black) together with the corresponding low-$q$ Fourier reconstruction (red dashed).

To directly visualize the local structure of the incommensurate crystal at the atomic scale, we performed STM measurements on a surface cleaved in ultrahigh vacuum in constant-current mode (details in the Methods). Based on the known bulk stacking sequence and the observed surface geometry, the exposed terrace is assigned to a $TaS_2$ layer. Fig. 3a shows an atomically resolved tunnelling-current map acquired at $V_s = 400$ mV and $I_s = 100$ pA. The tunnelling-current channel was acquired simultaneously with the $Z$-channel; we use the current channel here because the long-wavelength modulation is more clearly resolved in the current signal, while the corresponding $Z$-channel data is presented in the Supplementary Information. A pronounced long-wavelength modulation is visible across the field of view, in addition to the atomic-scale periodic structure. This modulation is more clearly resolved in the independently acquired smaller field-of-view image in Fig. 3b, obtained at $V_s = 400$mV and $I_s = 80$pA, where the long-wavelength variation is clearly superimposed on the atomic-scale contrast.

The Fourier transform of the STM current map shown in Fig. 3c reveals atomic-scale Fourier components at $|Q| \approx$ 3.3 cycles nm$^{-1}$, together with three distinct low-$q$ peaks at $|q| \approx 0.68$ cycles nm$^{-1}$. The low-$q$ peaks correspond to a characteristic real-space wavelength of approximately 1.4 nm and are oriented along directions distinct from those of the atomic-scale Fourier components. We refer to the higher-$Q$ features as atomic-scale Fourier components rather than crystallographic Bragg peaks, since the STM tunnelling signal is sensitive to the local electronic structure and tunnelling matrix elements and need not reproduce the crystallographic structure factor.[34] We therefore focus on the well-defined low-$q$ components, which directly capture the long-wavelength modulation, and use these components to reconstruct its real-space form. To isolate the long-wavelength component of the STM signal, we retained the experimentally observed low-$q$ Fourier components and performed an inverse Fourier transform. The resulting real-space reconstruction is shown in Fig. 3d. It recovers the long-wavelength modulation observed directly in the STM current maps, while suppressing the atomic-scale contrast. This correspondence between the modulation in the original STM image and the low-$q$ Fourier reconstruction demonstrates that the long-wavelength variation is associated with well-defined Fourier components rather than arising solely from a broad or slowly varying background in the image. The correspondence is further illustrated by the mean-subtracted line profiles shown in Fig. 3e, taken along the trajectory indicated in the STM image (Fig. 3a) and at the corresponding location in the low-$q$ reconstruction (Fig. 3d). The measured tunnelling-current profile (black) contains a slowly varying long-wavelength modulation on which the shorter-period atomic-scale variations are superimposed. The corresponding low-$q$ Fourier reconstruction (red dashed) follows the long-wavelength component of the experimental profile, providing a direct real-space correspondence between the modulation observed in the STM image and the low-$q$ Fourier components. Additional line profiles taken from spatially separated regions of the STM images are presented as Supplementary Information and show the same characteristic long-wavelength modulation, further demonstrating that the modulation is spatially widespread across the imaged surface.

The observed long-wavelength modulation is consistent with the incommensurate structural environment of the bulk crystal. The three low-$q$ peaks occur approximately at 0.68 cycles nm$^{-1}$, corresponding to a characteristic length scale of approximately 1.4 nm, and form a set of modulation vectors with orientations distinct from those of the atomic-scale Fourier components. The observed modulation is distinct from the conventional CDW of 2H-$TaS_2$ as STM measurements of 2H-$TaS_2$ show a quasi-$(3 \times 3)$ CDW whose modulation vectors are aligned with the underlying atomic reciprocal-lattice directions[35], whereas the low-$q$ features observed here have a different orientation and a different characteristic length scale. Taken together, these observations establish a well-defined, local long-wavelength modulation at the surface that is consistent with the incommensurate structural framework of the bulk crystal.

**Evidence of Superconductivity**

The temperature-dependent resistivity measured on two different single crystals from the same batch is shown in Fig. 4a. In the temperature range of 4-400 K, the resistivity decreases monotonically upon cooling, confirming the metallic nature of the samples. On further cooling, a pronounced two-step resistive transition is observed. At approximately 3.4 K, the resistivity begins to decrease sharply and reaches nearly 50% of its normal-state value around 3.0 K, below which another sharper resistive drop occurs with the resistivity becoming essentially zero at 2.3 K, establishing the occurrence of a zero-resistance superconducting state. The application of an external magnetic field systematically shifts both resistive anomalies toward lower temperatures (Fig. 4b-c), providing strong evidence for the superconducting origin of this transition. The observation of two distinct superconducting transitions suggests the presence of more than one superconducting component or superconducting energy scale within the crystal. Since the material possesses a layered/superlattice structure, the two transitions may arise from superconductivity developing in electronically inequivalent regions of the structure. One possible interpretation is that the higher-temperature transition corresponds to superconductivity associated predominantly with the individual conducting layers (intralayer superconductivity), whereas the lower-temperature transition may be associated with the establishment of coherent superconducting coupling between adjacent layers (interlayer superconductivity). In such a scenario, superconducting correlations could develop within the layers at a relatively higher temperature, while three-dimensional phase coherence mediated by interlayer coupling is established only at a lower temperature, producing the successive feature in resistivity and ultimately the zero-resistance state. Similar successive superconducting transitions have been reported in related members of the $ATa_2S_5$ family (A = Ba, Sr, Mg)[36–38]. In particular, a hierarchy of superconducting transitions in $SrTa_2S_5$ has been discussed in terms of distinct interlayer and intralayer superconducting responses[37]. The similarity between these systems suggests that the two-step transition observed here may likewise be related to the highly anisotropic and layered nature of the superconducting state. However, the assignment of the individual resistive anomalies specifically to intralayer and interlayer superconductivity should be regarded as a possible interpretation rather than definitive evidence based solely on resistivity measurements.

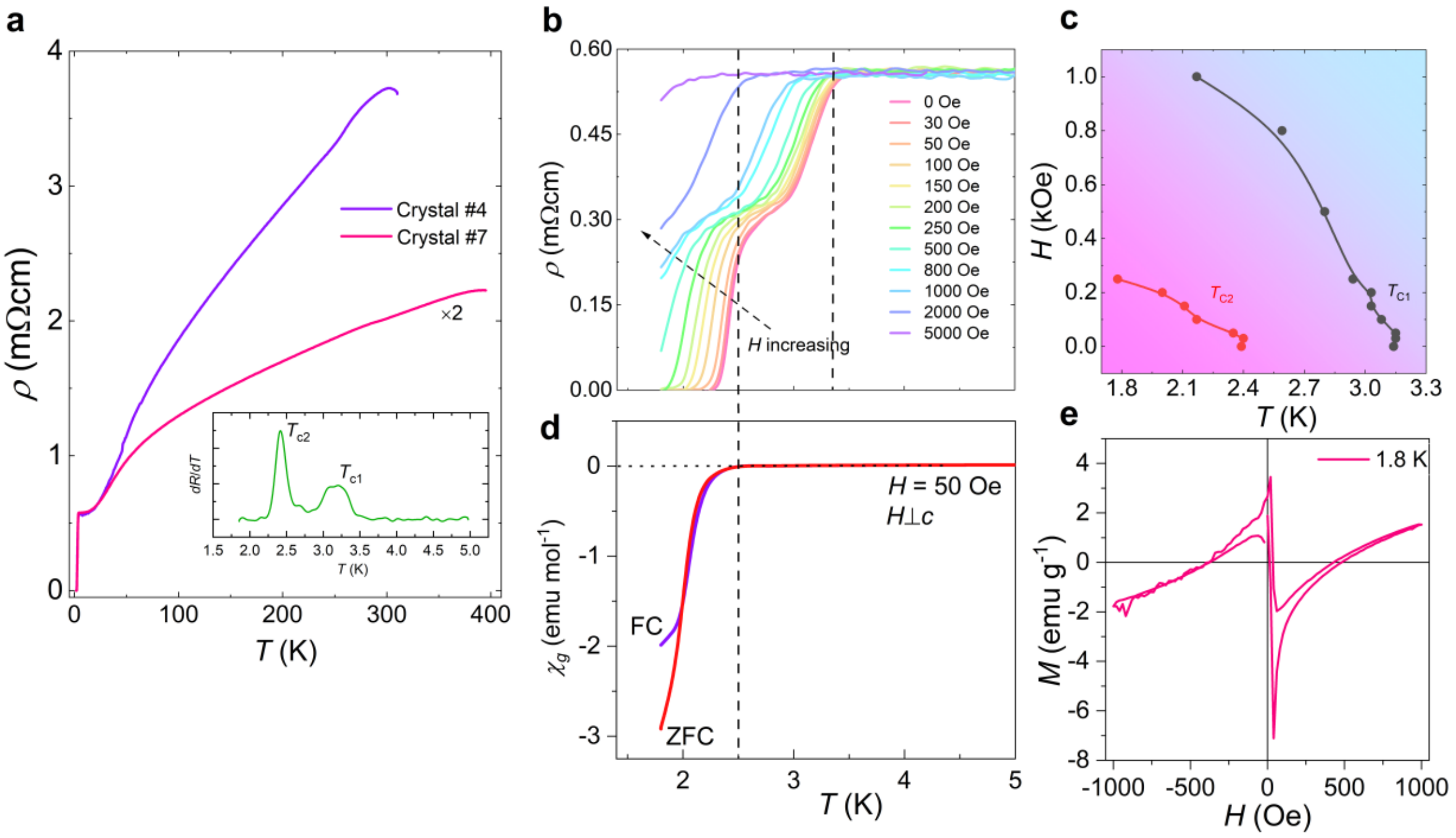

**Fig. 4| Electrical transport data on single crystals of $(Sr_6TaS_8)_{1+\delta}(TaS_2)_8$. a,** $\rho(T)$ data of two crystals in the temperature range of 1.8-400 K. **b,** Field dependent $\rho(T)$ data near the superconducting transition depicting the variation of both the transitions with applied magnetic field. **c,** The critical temperatures $T_{c1}$ and $T_{c2}$ for both transitions extracted from **b** as a function of applied field. **d,** Variation of superconducting transitions with applied magnetic field. Criteria of selection of $T_c$ is defined in the inset of **a**.

The superconducting transition is further corroborated by the magnetic susceptibility measurements performed on a single crystal (Fig. 4d). Both field-cooled (FC) and zero-field-cooled (ZFC) susceptibility measurements under a small applied magnetic field exhibit a pronounced diamagnetic response below approximately 2.5 K, coinciding with the lower-temperature resistive transition observed in the electrical transport measurements. The concurrence of the diamagnetic onset with the development of zero electrical resistance provides independent evidence for the bulk superconducting state. With increasing applied magnetic field, the superconducting transition shifts systematically toward lower temperatures, as expected for a type-II superconductor and consistent with the field dependence observed in the resistivity measurements (Supplementary Information).

Because the precise crystal structure and hence the crystallographic density of the sample are not yet established, an accurate conversion of the measured magnetic susceptibility into an absolute superconducting volume fraction is not possible. Nevertheless, the clear diamagnetic response in the ZFC measurement provides qualitative evidence for superconducting shielding. The difference between the FC and ZFC responses can additionally be associated with magnetic flux trapping and vortex pinning in the mixed state. Crystals from other batches also show a sharp diamagnetic transition below 2.5 K (Supplementary Information).

The field-dependent magnetization, $M(H)$, measured at 1.8 K (Fig. 4e) exhibits the characteristic hysteretic behaviour expected for a type-II superconductor. The finite width of the hysteresis loop indicates irreversible magnetization arising from the pinning of magnetic vortices, demonstrating that the sample enters a mixed state in an applied magnetic field. From the low-field deviation of the $M(H)$ curve from the initial linear Meissner response, the lower critical field (defined as $H_{c1}^{M}$) is estimated to be 25 Oe. The magnetization remains irreversible up to fields of approximately 400 Oe, providing an estimate of the upper critical field ($H_{c2}^{M}$) at 1.8 K. The relatively small values of both critical fields are consistent with the low superconducting transition temperature and indicate that superconductivity is readily suppressed by comparatively weak magnetic fields.

## Anisotropy of superconducting properties

To investigate the superconducting behavior further, field- and angle-dependent resistivity ($\rho$) measurements were conducted. As shown in Fig. 5a**,** for applied field in the range of $\pm 8$ kOe perpendicular to the *ab*-plane, the magnetoresistance show symmetrical V-shaped curve centered at zero field, which is due to pronounced suppression of superconducting transport with applied field. Two slope changes are observed as function of applied field. With increasing temperature from $1.8$ K to $10.0$ K, both these dips gradually broaden and shift toward higher resistivity, eventually getting fully suppressed, as indicated by black dotted arrow.

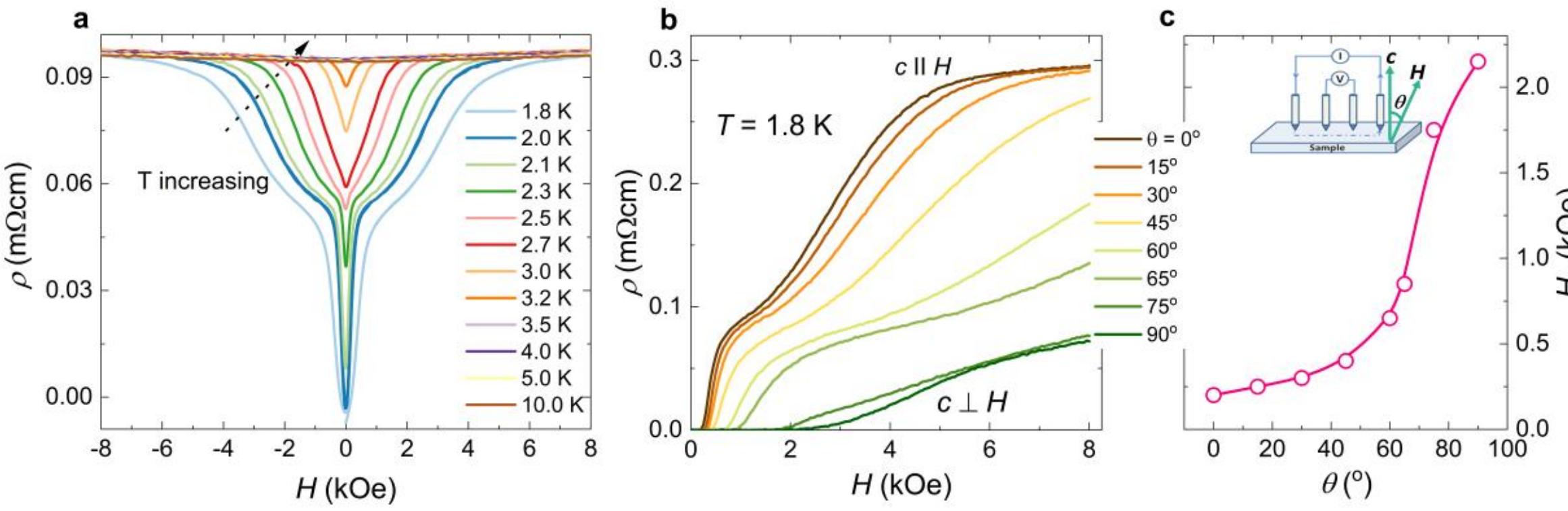

**Fig. 5| Magnetoresistance data on a single crystal of $(Sr_6TaS_8)_{1+\delta}(TaS_2)_8$. a,** Resistivity as a function of magnetic field ($H$) measured at various temperatures from $1.8$ K to $10.0$ K. The black dashed arrow indicates the direction of increasing temperature. **b,** Field-dependent resistivity data collected at $T = 1.8$ K for different tilt angles ($\theta$) between the crystallographic $c$-axis and the applied magnetic field ($H$), evolving from $c \parallel H$ ($\theta = 0°$) to $c \perp H$ ($\theta = 90°$). **c,** Angular dependence of the critical magnetic field $H_c(\theta)$ measured at $1.8$ K. The inset of **c** shows a schematic setup of the four-probe electrical transport measurement showing the definition of the angle $\theta$ relative to the magnetic field $H$ and the $c$-axis.

The anisotropy of the magneto-transport was investigated by varying the angle $\theta$ between the applied magnetic field ($H$) and the crystallographic $c$-axis at $T = 1.8$ K (Fig. 5b), within the low field range of interest i.e. $H$ < 8 kOe. For the configuration where the field is parallel to the $c$-axis ($\theta = 0°$, $c \parallel H$), the system transitions from superconducting to the resistive state (field is defined as $H_c^R$) at a significantly lower critical field ($H_c^R$ ~ 250 Oe) compared to the perpendicular configuration ($H_c^R$~ 2300 Oe) ($\theta = 90°$, $c \perp H$). In the intermediate orientations, as $\theta$ increases from $0°$ to $90°$, the onset of electrical resistivity systematically shifts toward higher magnetic field strengths. It is also observed that, the modulation in slope at higher field values also varies significantly with angle, and shifts to higher field values. The resulting angular phase boundary $H_c^R(\theta)$ at 1.8 K is plotted in Fig. 5c. The critical field displays a smooth, non-linear enhancement as $\theta$ approaches $90°$. This strong angular dependence highlights significant transport anisotropy in the system and the anisotropic 2D nature of superconductivity, attributed to the layered or quasi-2D nature of the material.

## Thermodynamic evidence for bulk superconductivity

The thermodynamic properties of the superconducting state were further investigated using low-temperature specific heat measurements. Fig. 6a presents the temperature dependence of the specific heat, $C_p(T)$, measured in zero magnetic field. The specific heat increases continuously with increasing temperature; however, a change in its temperature dependence is discernible around ($T_c$ ~ 3 K), as indicated by the vertical dashed line. More clearly, the $C_p/T$ versus $T^2$ plot (Fig. 6b) exhibits a broad but prominent feature in the vicinity of $T_c$, with separate linear regimes above and below the transition. This feature provides thermodynamic evidence for the superconducting transition identified from the resistivity and magnetic susceptibility measurements. Another subtle change in the slope of $C_p/T$ versus $T^2$ plot is observed at slightly higher temperature ~3.5 K corresponding to $T_{c1}$ as observed in the resistivity measurements (inset of Fig. 6b).

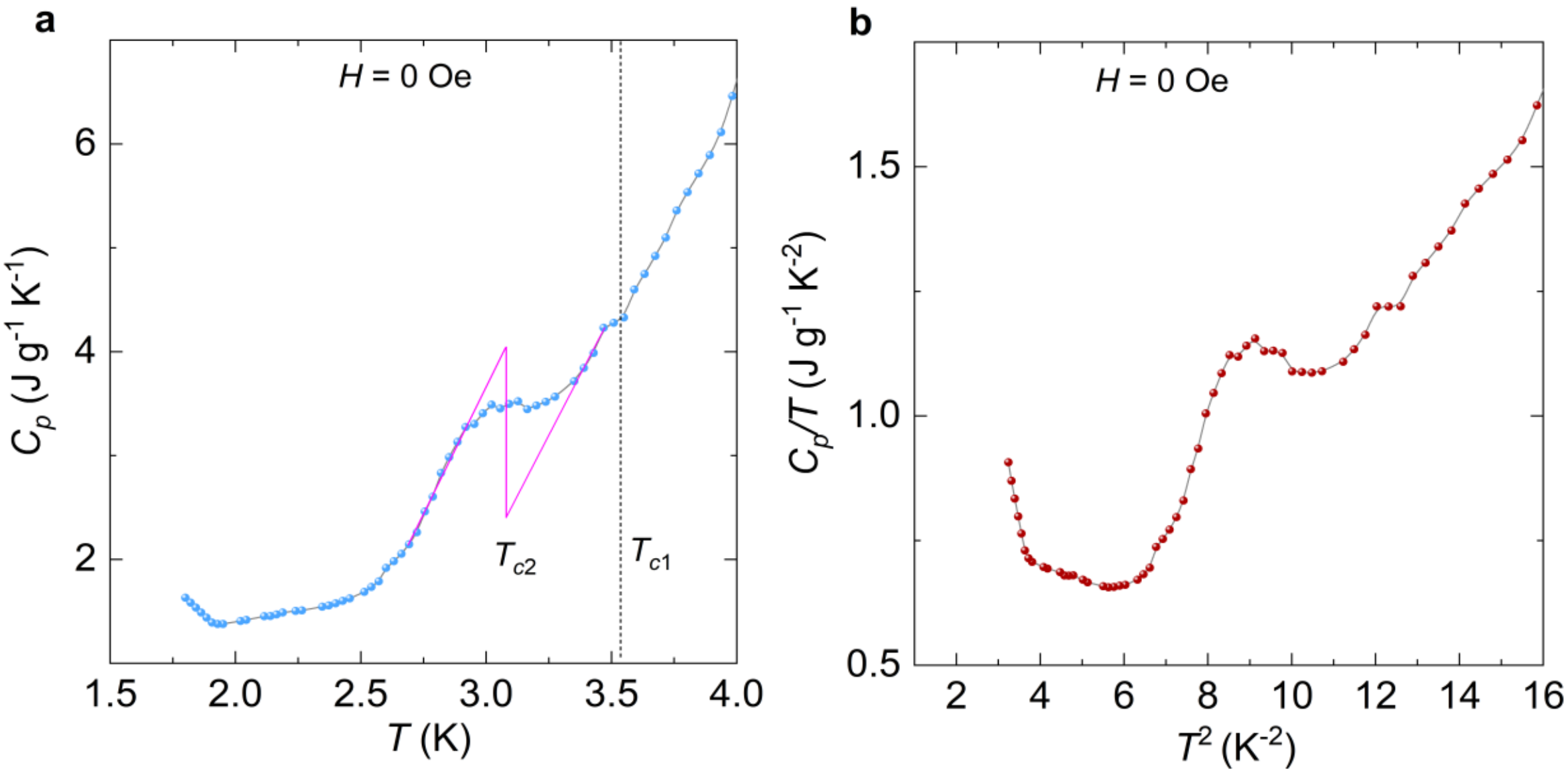

**Fig. 6| Heat capacity data of a single crystal**. **a**, $C_p(T)$ curve at zero applied magnetic field indicating a weak anomaly at 3 K ($T_{c2}$). **b,** $C_p/T$ vs $T^2$ curve better demonstrate the anomaly associated with superconducting transition termed as $T_{c1}$ corresponding to resistivity measurements.

In a homogeneous bulk superconductor undergoing a sharp second-order superconducting transition, a pronounced specific-heat jump is generally expected at $T_c$. In the present case, however, the feature is relatively broadened. Such broadening may originate from a distribution of local superconducting transition temperatures arising from compositional variations, structural disorder, strain, or microscopic inhomogeneity within the crystal. In addition, the presence of multiple superconducting components or weakly coupled superconducting regions, as suggested by the two-step resistive transition, could contribute to a broadened thermodynamic signature rather than a single well-defined specific-heat jump.

Interestingly, the zero-field $C_p(T)$ data also exhibit an upturn at the lowest measured temperatures, suggesting the possibility of an additional low-temperature feature. While this feature may be associated with another phase transition or an additional superconducting component, its origin cannot be established unambiguously from the present specific-heat data alone. Possible contributions from magnetic excitations, Schottky-type anomalies, nuclear contributions, or other low-energy degrees of freedom should also be considered. Measurements under applied magnetic fields, together with higher-resolution specific-heat data extending to lower temperatures, would be useful for determining whether this low-temperature feature is intrinsically associated with superconductivity. The relatively weak and broadened feature, however, suggests that the superconducting state may be affected by sample inhomogeneity, disorder, or multiple superconducting energy scales.

Taken together, the resistivity, magnetic susceptibility, magnetization and heat capacity measurements provide mutually consistent evidence for superconductivity below approximately 2.5 K. In particular, the field-induced suppression of the resistive and diamagnetic transitions, together with the $M(H)$ response and finite critical fields, establishes the type-II nature of the superconducting state. The magnetic and heat capacity measurements provide independent support for the superconducting transition inferred from transport measurements.

## Conclusion

We reproducibly observe superconductivity below $T_c$ ~ 2.5 K in the intrinsically synthesized bulk moiré compound $(Sr_6TaS_8)_{1+\delta}(TaS_2)_8$, as evidenced by complementary transport, magnetic susceptibility, magnetization, and low-temperature specific-heat measurements. The superconducting transition is consistently observed across multiple independently prepared growth batches and is also present in polycrystalline samples, demonstrating the reproducibility of the superconducting state. Additionally, we provide direct evidence of long-wavelength modulations due to incommensuration leading to moiré patterns through scanning tunnelling microscopy. The observation of superconductivity in an intrinsically synthesized bulk moiré crystal provides an opportunity to investigate the interplay between moiré-induced electronic reconstruction and superconductivity in a three-dimensional bulk material. Further studies of the field dependence, anisotropy, electronic structure, and structural response will be important for elucidating the microscopic mechanisms governing superconductivity. It would also be interesting to see if superconductivity can persist in few layers of the materials. If realised, it will open an easier channel for fabrication of thin-film superconducting devices.

**Methods:**

Sample growth: Single crystal samples of $(Sr_6TaS_8)_{1+d}(TaS_2)_8$ Moiré system were synthesized using a general strategy reported in reference.[15] SrS (Alfa Aesar, 99.9%), Ta powder (Alfa Aesar, 99.97%), S powder (Sigma-Aldrich, 99.9%), and anhydrous $SrCl_2$ (Sigma-Aldrich, 99.9%) were used without further purification. All procedures were carried out in an argon-filled glove box. The growth process was optimised in three batches by adjusting the reaction profile. Precursors were mixed, finely ground, and placed in an alumina crucible. This crucible was then enclosed inside a quartz ampoule (15 mm ID × 18 mm OD), evacuated to $10^{-5}$ Torr, and flame-sealed. The ampoule, ~ 12 cm long, was positioned upright in a programmable muffle furnace. After the heating profile, black shiny crystals were found to be embedded inside yellowish powdered mass which were harvested manually. Many thin crystals were also found sticking to the outer and inner walls of the crucible which could be lifted using a Kapton or scotch tape. The detailed reaction conditions are presented in Table 1. It was observed that the crystals from growth attempt by slow cooling from 950 ºC (Batch #1 and #2) contained trace amount of chlorine. However, crystals grown at a higher temperature (1050 ºC) contained only negligible or no chlorine. The crystals from batch #4 should yield STS(11,88º) as reported in the previous work.[15]

**Table 1.** Detailed synthesis conditions for all three batches of the sample.

| Batch | Precursors | Ratio | Temperature Profile | Product |
|---|---|---|---|---|
| 1 | SrS, Ta, S, $SrCl_2$ | 4:4:12:1 | 24 h ramp to 950°C; 48 h soaking; cooling to 750°C in 120 h; cooling to RT in 24 h | Black irregular/hexagonal plates with yellowish solid |
| 2 | SrS, Ta, S, $SrCl_2$ | 4:4:12:1 | 24 h ramp to 950°C; 48 h soaking; cooling to 750°C in 168 h; cooling to RT in 24 h | Black irregular/hexagonal plates with yellowish solid |
| 4 | SrS, Ta, S, $SrCl_2$ | 4:4:12:1 | 24 h ramp to 750°C; 24 h soaking; 5 h ramp to 1050°C; 100 h soaking; cooling to 750°C in 100 h; cooling to RT in 24 h | Black irregular/hexagonal/ elongated plates with yellowish solid |

Single Crystal X-ray Diffraction: A suitable single crystal was mounted on the goniometer head using a MiTeGen polymer loop. Single-crystal X-ray diffraction data were collected on a Rigaku XtaLAB Synergy-S microfocus diffractometer equipped with a HyPix-6000HE hybrid photon-counting detector. Mo $K_\alpha$ radiation (λ = 0.71073 Å) was generated using a PhotonJet microfocus X-ray source operated at 50 kV and 1 mA. The data were collected in the step size of 0.5° with an exposure time of 30 s per frame, giving a total of 1420 frames. The diffraction data were processed using the CrysAlis PRO software suite (Rigaku Oxford Diffraction).[39] The precession images (Figure x, y) in reciprocal space were constructed with unit cell parameters: 3.32, 3.32, 12.42, 90°, 90°, 120°.

FIB Sample preparation: A thin cross-section lamella of the $(Sr_6TaS_8)_{1+\delta}(TaS_2)_8$ sample was prepared using a dual-beam FIB-SEM Scios microscope (ThermoFisher Scientific) equipped with a Ga ion source. A few polycrystalline powders were dispersed in methanol solutions and then sonicated for a few minutes. The resultant suspension was drop-cast onto a Si(100) substrate. A Pt protective layer with a thickness of approximately 1.5 µm was subsequently deposited onto the selected area of the small crystal to protect them from the Ga ion irradiation.

A lamella with approximately dimensions of 6 µm x 3 µm x 6 µm was successfully prepared by FIB milling and detached from the substrate using a U-shaped undercut. The lamella was subsequently lifted out using an Omniprobe micromanipulator and mounted onto a TEM semi-grid. Final thinning and polishing were performed at an ion beam energy of 5 kV, with the beam current progressively reduced from 2.8 nA to 50 pA. resulting in a final lamella thickness of around ~60 nm. The prepared lamella was subsequently characterised using a Titan-Themis transmission electron microscope.

Transmission Electron Microscopy: High-resolution electron microscopy characterisations were performed using an aberration-corrected ThermoFisher Titan Themis microscope. To achieve optimal spatial resolution, the instrument is equipped with a high-brightness field emission gun (X-FEG), and a CEOS probe spherical aberration ($C_s$) corrector. The microscopes offer spatial resolution of around 60 pm. High-resolution (S)TEM imaging were conducted using high-angle annular dark field (HAADF), detectors. HAADF-STEM images were recorded with a convergence semi-angle of 24.5 mrad, a probe current of 85 pA, and an inner collection angle of 50 mrad. Velox software is used to map the intensity profile and to process and analyse the data.

Scanning Electron Microscopy: FESEM images were collected from different crystals using a Thermoscientific Apreo 2S microscope equipped with a field emission gun and various detectors. The samples were coated with Au using a Quorum Q150TS plus coater. The Energy dispersive X-ray analysis (EDX) was performed on various points on the different crystals using APEX software and an ETD detector for compositional analysis.

Scanning Tunnelling Microscopy: STM experiments were performed using a low temperature ultrahigh vacuum (UHV) STM (Unisoku 1300J) at 77K. The sample was cooled to 77K and cleaved in-situ UHV at a pressure better than $10^{-10}$ mbar. The cleaved sample was immediately transferred to STM head at 77K. Images are acquired by cut Pt-Ir tips. Reproducibility of the features was ensured by scanning spatially distinct regions of the sample with different in-situ changed tips by fast scanning and pulsing, all of which retained the features discussed in this manuscript.

Physical Characterization methods: DC susceptibility data were collected using a SQUID magnetometer (Quantum Design MPMS 3) in an applied magnetic field of 50-1000 Oe in both zero field cooled (ZFC) and field cooled (FC) protocols. Isothermal magnetisation data was collected at different temperatures in maximum field of $\pm$1000 Oe. The data were recorded in warming cycles.

DC resistance of small crystals was measured with and without applied magnetic field using a PPMS (Quantum Design Dynacool). Conducting silver paste was used to make contacts on the sample using a 100 µm thick copper wire.

Heat capacity was measured on a small plate shaped crystal with a mass ~ 5 mg using relaxation method in a heat capacity option of PPMS (Quantum Design Dynacool). Addenda was subtracted from the sample data.

**Acknowledgements:**

GST and PM acknowledge ANRF (Govt. of India) for the Early Career Research Grant (sanction no.: ANRF/ECRG/2024/ 001436/CS) for the financial support. GST and RS thank IISER Berhampur for the generous financial support in the form of a seed grant and SN thanks IISER Berhampur for a postdoctoral fellowship. The authors from IISER Berhampur thank the Director, IISER Berhampur, for the extended support in lab establishment. Central Advanced Instrumentation Facility (CAIF) at IISERBPR is acknowledged for providing instrumental support.

**Author Contribution:** SKN and PM: synthesis and basic characterisation, SM and SMB: HAADF-STEM measurements, SK and PPJ: single crystal X-ray diffraction studies, PKM: Transport, magnetic and heat capacity data collection, SC: STM imaging and data analysis, RS: conceptualization, STM imaging, data analysis, and manuscript writing, GST: conceptualization, funding acquisition, data analysis, plotting and manuscript writing. All the authors contributed to manuscript writing and reviewing.

**Corresponding Author**: gsthakur@iiserbpr.ac.in; rahulsharma@iiserbpr.ac.in

**Supplementary Information**

## 1. Compositional Analysis

We first establish the identity of the grown samples using SEM-EDX. Multiple crystals from each batch were screened at several points on the same crystal. From SEM-EDX, the elemental ratio in the crystals was found close to 1:1.3:3.6 for Sr:Ta:S with ~10 % deviation from one batch to another, which is consistent with the previous report (figure 1)[1] and with the expected elemental ratio considering the composition $(Sr_6TaS_8)_{1+\delta}(TaS_2)_8$ with $\delta = 0.1$. EDX mapping shows the presence of all the elements in the crystal.

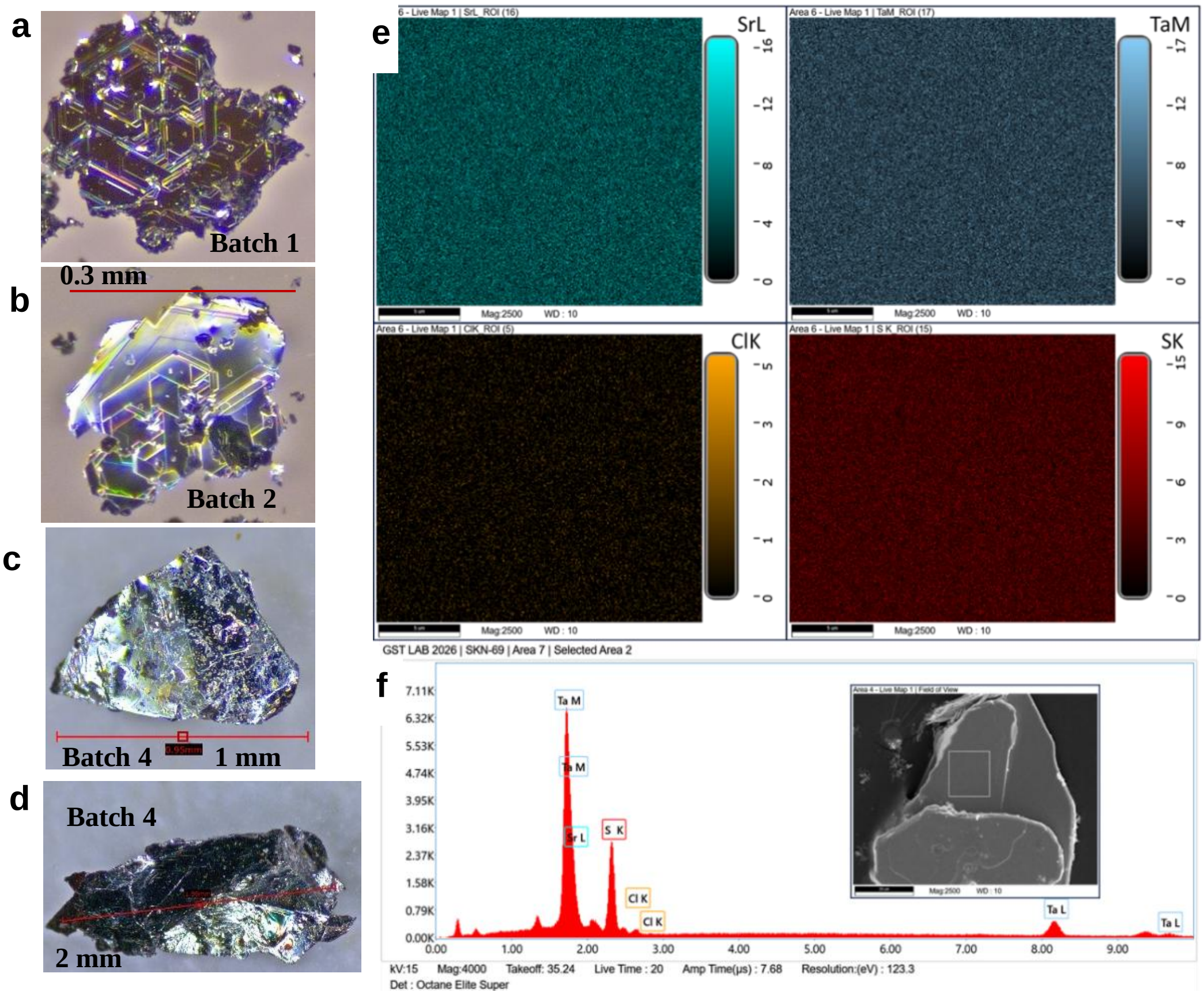

g

| Element | Atomic % | | | | | |
|---|---|---|---|---|---|---|
| | Crystal 1 | Crystal 2 | Crystal 3 | Crystal 4 | Crystal 5 | Crystal 6 |
| **S** | 53.1 | 58 | 59.3 | 58.3 | 59.5 | 59.6 |
| **Cl** | 0.1 | 0.9 | 0.6 | 0.7 | 0.6 | 0.8 |
| **Sr** | 16.9 | 17.5 | 18 | 18 | 17.9 | 18 |
| **Ta** | 29.9 | 23.6 | 22.1 | 22.9 | 21.9 | 21.6 |

**FIG. 1. Compositional analysis using SEM-EDX. a-d**, optical images of few representive crystals from different batches. **e,** EDX mapping on a crystal. **g**, its SEM spectra and and compositional analysis of different crystals by EDX.

2. Single Crystal X-ray Diffraction.

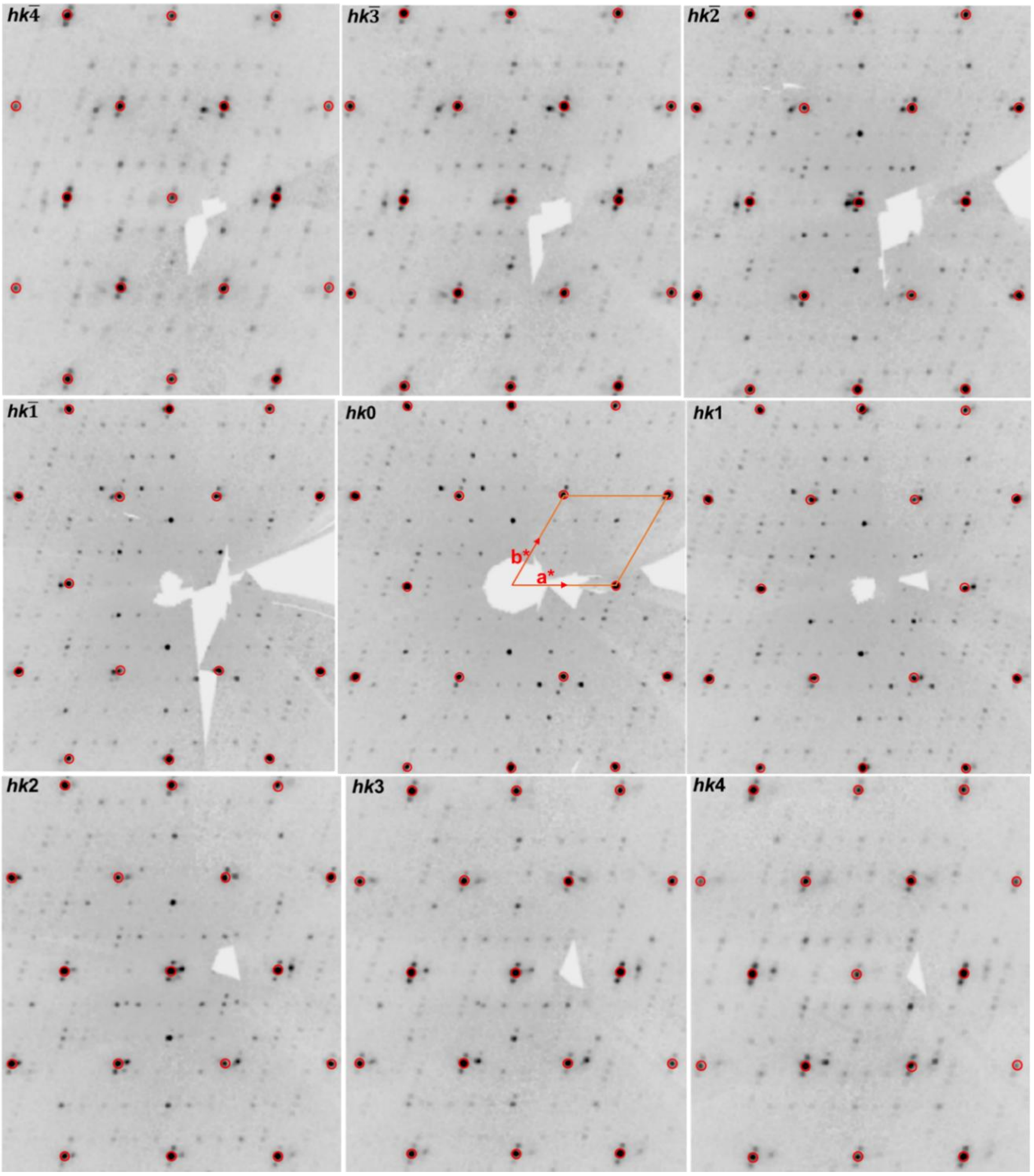

**FIG. 2. Single Crystal X-ray Diffraction.** Reconstructed $hkl$ reciprocal-space sections ($l = -4$ to $+4$) obtained from single-crystal X-ray diffraction. The strong reflections marked by red circles correspond to the hexagonal $TaS_2$ sublattice, while weaker additional reflections arise from the incommensurate Moiré superlattice. The systematic appearance of these satellite reflections around the fundamental Bragg peaks provides evidence for a coherent long-range Moiré modulation.

## 3. HAADF STEM and Electron Diffraction

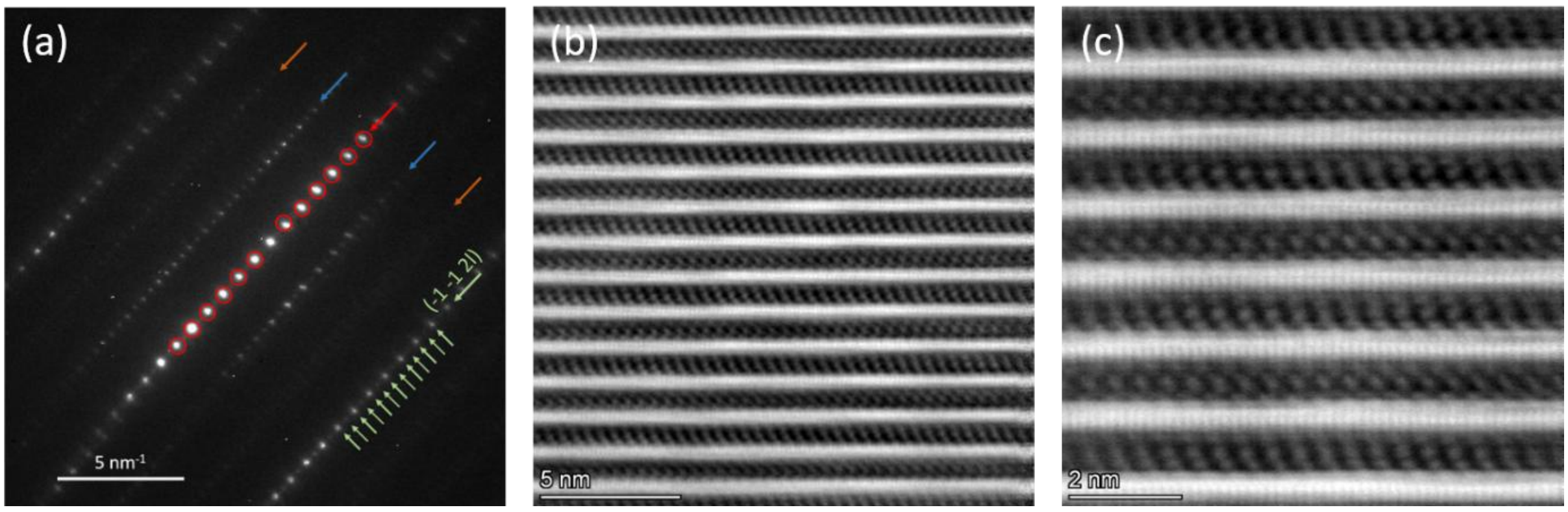

**FIG. 3. Atomic resolution scanning transmission electron micrograph of $(Sr_6TaS_8)_{1+\delta}(TaS_2)_8$. a,** SAED pattern of $(Sr_6TaS_8)_{1+\delta}(TaS_2)_8$ along a $TaS_2$ <120> type zone axis. The red circle marks the (000l) reflection along the *c* axis. Blue and orange arrows mark the rows of reflections near one-third and two-thirds of the $TaS_2$ $\{11\bar{2}0\}$ reciprocal spacing. **b,** HAADF-STEM image of a FIB-prepared $(Sr_6TaS_8)_{1+\delta}(TaS_2)_8$ crystal along the [010] zone axis, showing the alternating stacking of the bright $TaS_2$ and dark $Sr_6TaS_8$ layers over a large area. **c**, Raw (unprocessed, grey-scale) HAADF STEM image corresponding to the colour image Fig. 2c in main text.

## 4. Scanning Tunnelling Microscopy.

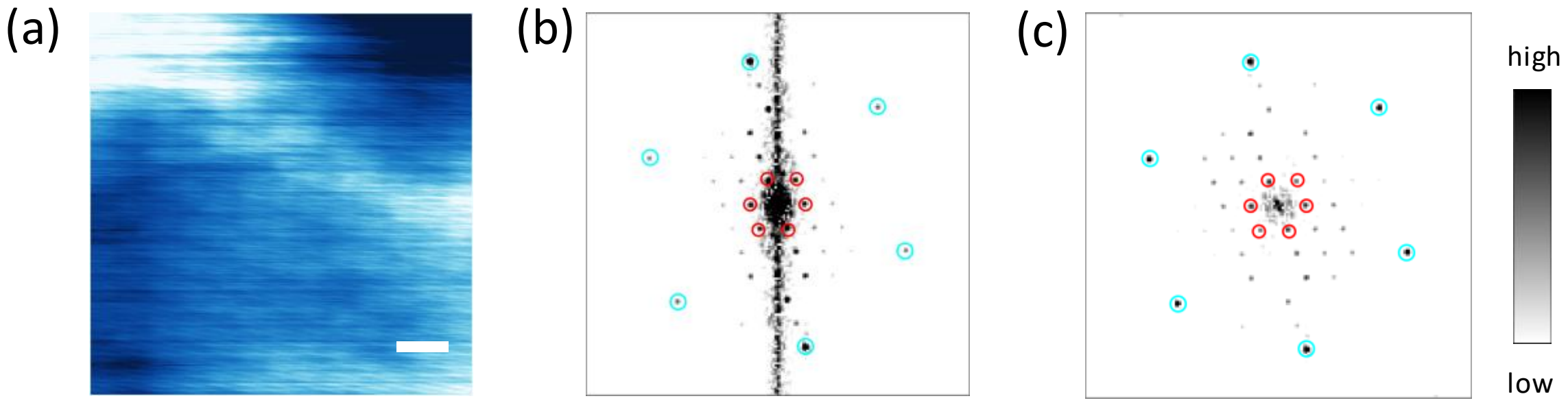

**FIG. 4. Consistency in $\boldsymbol{Z(r)}$ and $\boldsymbol{I(r)}$. a,** $Z(r)$ acquired at $V_S = 400$ mV and $I_S = 100$ pA, simultaneously with the $I(r)$ shown in the Fig. 3a of the main text. **b,** Fourier transform of the $Z(r)$ map with cyan and red circles at the identical positons as in Fig. 3c of main text showing that the features are reproduced reliably in $Z$ map too. However there is a scan related artefact which shows up as a vertical line and this is why we chose $I(r)$ for the main-text. **c,** Fig. 3c from the main-text, reproduced here to compare and show all Fourier transform features discussed in main text for $I(r)$ are replaced in $Z(r)$.

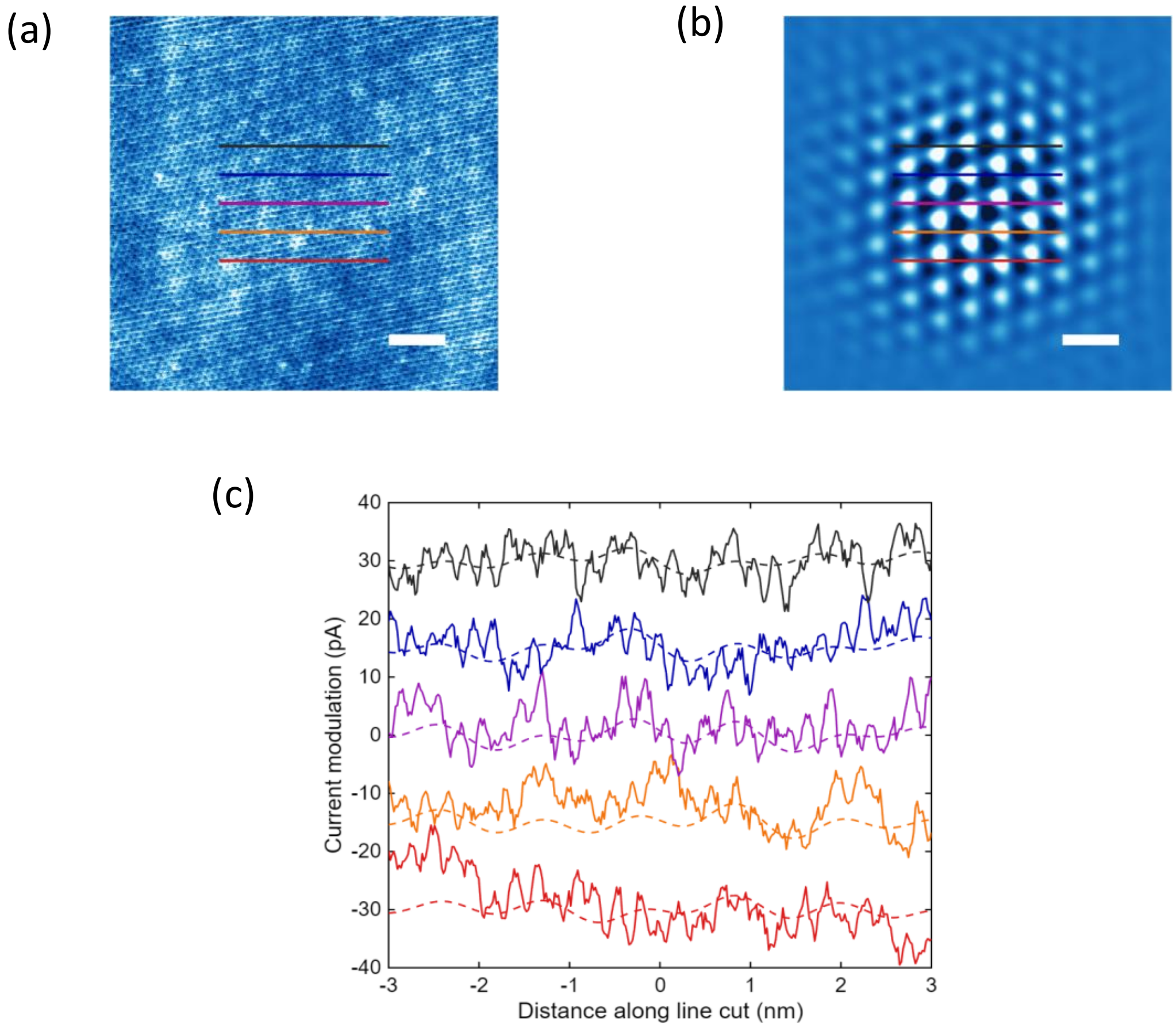

**FIG. 5. Line profiles showing the low-q modulation. a**, Locations where line profiles are taken on $I(r)$. **b,** Identical locations in inverse Fourier transform of low-$q$ modulations identical to Fig. 3 d. **c**, Line profiles offset by 15 pA. The measured tunnelling-current profile (solid) contains a slowly varying long-wavelength modulation on which the shorter-period atomic-scale variations are superimposed. The corresponding low-$q$ Fourier reconstruction (dashed) follows the long-wavelength component of the experimental profile, providing a direct real-space correspondence between the modulation observed in the STM image and the low-$q$ Fourier components.

5. Transport data and Magnetic data

Few crystals from other batches were also tested for superconductivity by both resistance and magnetic susceptibility measurements. Smaller crystals from batch 1 and 2 consistently showed sharp drop in resistivity at 2.5 K (Fig. 6). However, resistance measured with 2-point probe on small crystals did not drop to zero due to large contribution from contact and lead resistance (Fig. 6. a-b). Nevertheless, the field dependence of the resistance was reproduced in each measured crystal indicating that the drop in resistance is superconducting in origin. This is also corroborated by magnetic susceptibility measurements. In some crystals we observed an anomalous bump or shoulder between 240-300 K in the resistance data, however, its occurance and the thermal hystereis could not be reproduced reliably in all crystals (Fig. 6. c).

Magnetic suceptibility data collected on different batches of crystals is shown in Fig. 7. Measurement were made on two different set of samples from each batch. In one set, crystals sticking on the surface of the alumina crucible were exfoliated onto the kapton tape and put in the magnetometer (bunch of crystals on tape). In other set, bunch of small crystals manually picked from the grown mass was measured (bulk crystals). All the samples from different batches show supeconducting transition to diamagnetic state below 2.5 K. In the measurement on a single crystal additional broad transition at higher tempertaure (below 4 K) is observed which is consitent with the resistivity and heat capacity data on the crystal of same batch presented in the main text (Fig. 4b and 6a-b).

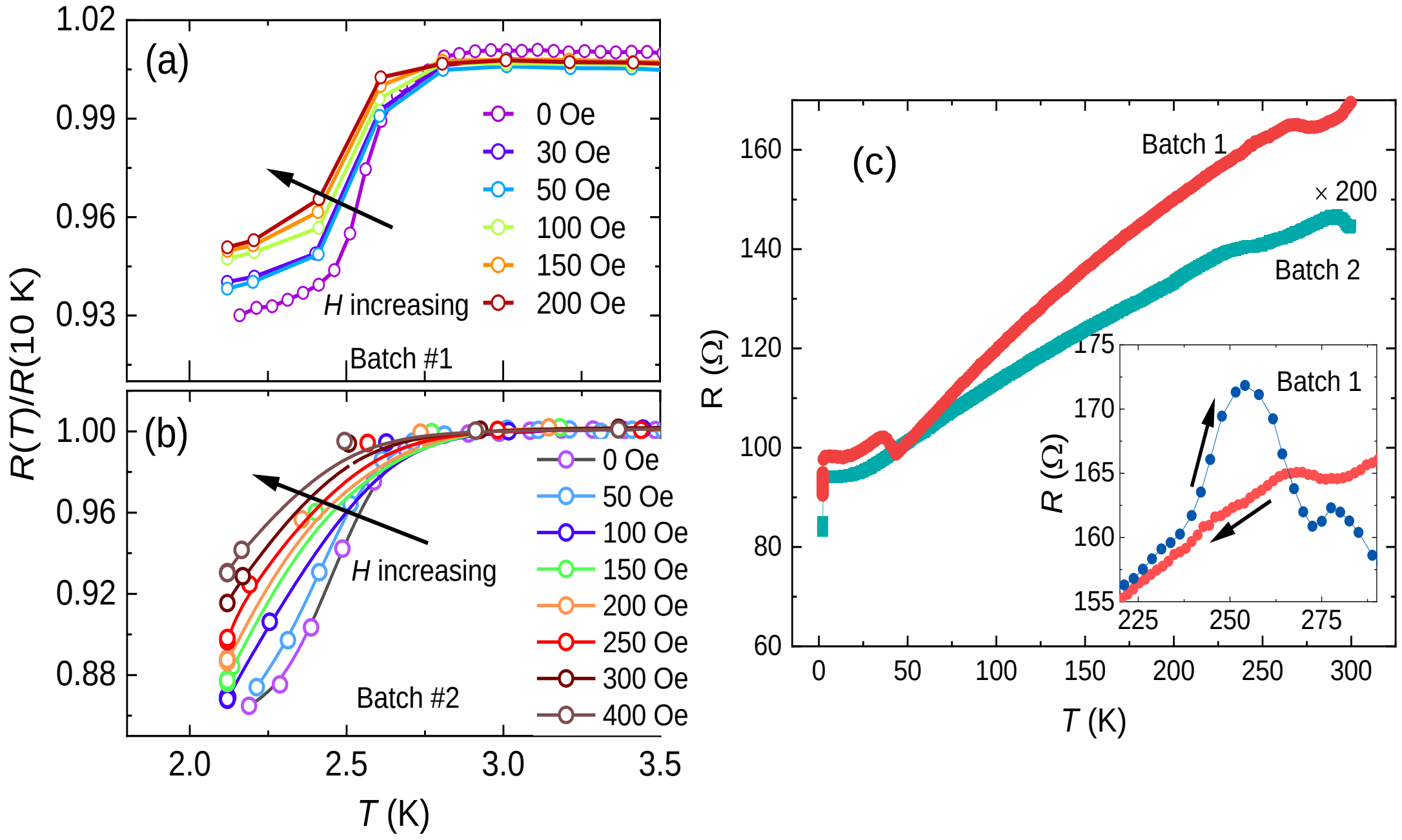

**FIG 6. Consistency of superconducting transition in crystals grown in different batches. a-b,** Resistance versus temperature data of two single crystals from different batch measured using 2-point point probe method in a PPMS. The resistance do not drops to zero due to large contribution from the contacts and lead. **c,** Resistance data in temperature range of 2-310 K. An anomaly beginning at 250 K and above is observed however it was not reproducible in other crystals.

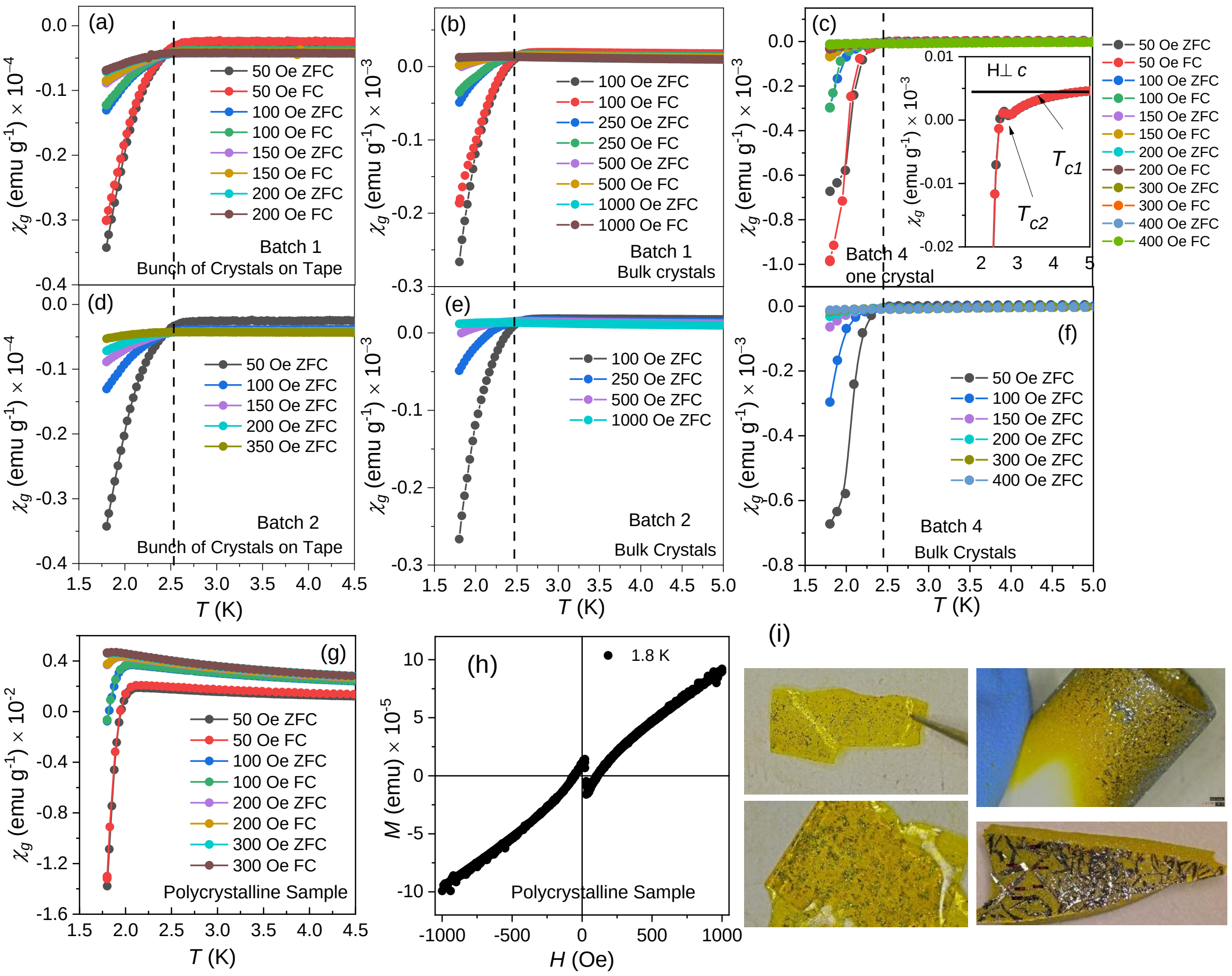

**FIG. 7. Susceptibility data in low temperature range near superconducting transition. a and d,** Data collected on crystals harvested from the outer side of the crucible on a piece of Kapton tape. **b and e**, Data collected on a bunch of loose crystals in random orientation. (c) Susceptibility data collected on one crystal with $H \perp c$. **f,** Magnified view of the data in **c** showing a broad transition beginning below 4 K and a sharp transition at 2.5 K. **g**, Susceptibility data collected on sintered polycrystalline pellet. **h**, M(H) hysteresis loop collected at 1.8 K on a polycrystalline sample. Pictures of crystals exfoliated on the Kapton tape from the out and inner surface of the alumina crucible.